\documentclass[%
 reprint,
superscriptaddress,
bibnotes,
 amsmath,amssymb,
 aps,
]{revtex4-2}

\usepackage[utf8]{inputenc}
\usepackage{siunitx}
\usepackage{array}
\usepackage{physics}
\usepackage{dsfont}
\usepackage{booktabs}
\usepackage[english]{babel}			
\usepackage{xcolor}
\usepackage{graphicx}				
\usepackage{amssymb,amsmath}		
\usepackage{url}					
\usepackage{marvosym}				
\usepackage[T1]{fontenc}            %
\usepackage{wrapfig}				
\usepackage{charter} 				
\usepackage{blindtext}				
\usepackage{datetime}				
\usepackage{lipsum}                 
\usepackage{float}                  
\usepackage{makecell, tabularx}
\usepackage{fancybox}  
\usepackage{dcolumn}
\usepackage{bm}
\usepackage{hyperref}
\hypersetup{colorlinks=true,allcolors=blue}

\newcommand{\UIBK}{Institute f{\"u}r Experimentalphysik, Universit{\"a}t Innsbruck, 6020 Innsbruck, Austria}
\newcommand{\ACP}{Friedrich Schiller University Jena, Abbe Center of Photonics, Institute of Applied Physics, 07745 Jena, Germany}
\newcommand{\IOF}{Fraunhofer Institute for Applied Optics and Precision Engineering IOF, Center of Excellence in Photonics, 07745 Jena, Germany}

\preprint{APS/123-QED}

\begin{document}
\title{How to measure laser chirp rate at single-emitter excitation energies}

%

\author{Timothée Mounier}
\affiliation{\UIBK}
\author{Moritz Kaiser}
\affiliation{\UIBK}
\author{Mert Tuncel}
\affiliation{\UIBK}
\author{Iker Avila Arenas}
\affiliation{\UIBK}
\author{René Schwarz}
\affiliation{\UIBK}
\author{Ria G. Krämer}
\affiliation{\ACP}
\author{Stefan Nolte}
\affiliation{\ACP}
\affiliation{\IOF}
\author{Florian Kappe}
\affiliation{\UIBK}
\author{Yusuf Karli}
\affiliation{\UIBK}
\author{Gregor Weihs}
\affiliation{\UIBK}
\author{Vikas Remesh}
\affiliation{\UIBK}
\affiliation{Email:vikas.remesh@uibk.ac.at}

\date{\today}

\begin{abstract}
We present a simple and direct method for measuring laser chirp rate, i.e., group delay dispersion (GDD) of ultrashort laser pulses at power levels compatible with single-quantum-emitter excitation. Traditional pulse characterization techniques rely on nonlinear optical processes that require high peak powers, making them unsuitable for the attojoule-to-femtojoule regime relevant to quantum photonics. Our approach utilizes a wavelength-to-time mapping method in which the arrival times of spectrally filtered components of a broadband pulse are recorded using a superconducting nanowire single-photon detector and correlated via a high-resolution time-tagging system. The resulting linear relationship between wavelength and arrival time directly yields the dispersion parameter and, subsequently, the GDD. Beyond single-emitter excitation, this technique can be applied in areas such as single-photon spectroscopy, ultralow-power optical communications, and time-domain quantum control, where linear and non-destructive dispersion characterization is essential.
\end{abstract}

\maketitle
\section{introduction}
Chirped laser pulses have found immense applications in various fields of fundamental and applied research, for instance in coherent control of atomic and molecular energy transition pathways \cite{warren1993coherent,chatel_competition_2003}, manipulating nanoscale nonlinear optical responses \cite{remesh_coherent_2019}, efficient and robust excitation of quantum dots\cite{kappe2024chirped}, high power laser development \cite{backus_high_1998}, dispersion engineering in optical communications systems, to name a few. A chirped pulse is characterized by a time-dependent variation in instantaneous frequency, where the instantaneous frequency either increases (up-chirp) or decreases (down-chirp) over the duration of the pulse. For some applications such as imaging or signal transmission, chirp results in detrimental effects, while for some, for instance, excitation of a quantum system, chirp is a useful tool. In fiber-based optical communication links, propagation-induced dispersion broadens the pulses and degrades signal integrity. In semiconductor quantum dots (QDs), in contrast, chirped excitation enables adiabatic rapid passage (ARP), a robust population inversion technique that mitigates effects of spectral diffusion and power fluctuations \cite{ramachandran_experimental_2021,kappe2022collective}. This paves the way to create robust and field-deployable, frequency-multiplexed quantum light sources. In laser technology, chirping a short pulse before amplification and then recompressing it afterward allows for producing petawatt-scale laser pulses. In all these cases, characterization of chirp becomes critical to optimizing the system performance. 

To characterize laser pulses, several techniques have been developed, involving indirect reconstruction \cite{trebino_measuring_1997,iaconis1998spectral,stibenz2005interferometric,miranda2012characterization,Gabolde2006,DeLong1994,Geib2019,Miranda2016,Loriot2013}, or reconstruction and compensation \cite{lozovoy2004multiphoton,Comin2014} that rely on nonlinear optical processes \cite{Alonso2020,Birge2006,chilla1991direct} such as second harmonic generation, sum-frequency generation, or four-wave mixing in bulk crystals or nanostructures \cite{Accanto2014}. Another class of techniques involves direct, time-domain pulse sampling using sub-cycle gating processes such as attosecond streak cameras \cite{Itatani2002} or tunneling ionization \cite{Park2018}. In almost all of these cases, one needs picojoule to nanojoule pulse energies \cite{trebino_measuring_1997,stibenz2005interferometric,lozovoy2004multiphoton}. Some works \cite{Zhang2003,Zacharias2025} report pulse characterization at femtojoule levels, though they require higher energy gate pulses. In Ref. \cite{ranka1997autocorrelation,chong2014autocorrelation} the authors employed a two-photon absorbing photodiode-based autocorrelator to extract laser pulse properties at energies as low as few picojoule. The fringe-resolved interferometric autocorrelation trace is affected by chirp, however, the evaluation is not straightforward, but rather complex algorithms have to be employed \cite{kleimeier2010autocorrelation}. Furthermore, the two-photon absorption in the detector also suffers from low efficiency at very high dispersion. 

For single-emitter chirped excitation experiments, the employed laser pulses carry pulse energies as low as attojoule \cite{remesh2023compact}. In such scenarios, nonlinear optics-based dispersion characterization techniques are largely ineffective. In Ref. \cite{Kaldewey2017,Kaldewey2018,kappe2022collective} the authors employed an intensity autocorrelator to indirectly estimate the chirp rate upto the level where nonlinear signal was detectable, well below the employed chirp rate in the experiment. Note that the presence of chirp is not directly evident in an intensity autocorrelation measurement. In Ref. \cite{Gamouras2013}, the authors characterized the chirp rate by emulating the optical path to a quantum emitter using an additional microscope objective in a separate optical path, increasing the resource overhead of the system. In the future, where chip-integrated quantum photonic devices may dominate, there exists a need for dispersion characterization techniques that are compatible with integrated optics, and can work at very low pulse energies. 

Several methods of linear pulse characterization have also been reported in the literature \cite{lepetit1995linear,Fittinghoff1996,geindre2001single,sainz1994real,kovacs2005dispersion,Bowlan2007,dorrer2008linear,fan_measurement_2013,Davis2018,kurzyna2022variable}. They either 
rely on spectral or temporal interference, employ temporal phase modulators and electronic elements, in combination with Fourier transform phase retrieval algorithms, which limit their application in real-time dispersion analysis. 
Furthermore, the capabilities of such techniques have not been demonstrated for measuring very high chirps or extremely weak pulses. The main limitation of interferometric methods is their extremely rigorous spatio-temporal coherence demands, and the requirement of well-characterized reference pulses. 

In this work, we device and demonstrate a simple experimental method to measure large dispersions, at power levels of few attojoule and lower, directly in the time domain. Our technique relies on the wavelength-to-time mapping effect when laser pulses are dispersed, and does not rely on temporal phase modulators or sensitive interferometers, and works for any spectral range. This technique provides a scalable, noise-tolerant, and alignment-insensitive pathway for quantifying chirp and time-bandwidth product in quantum photonic systems, enabling dispersion metrology at the energy scale relevant for single quantum-dot excitation and single-photon-level ultrafast optics.

\section{Concept}
\begin{figure}[htb!]
	\centering
	\includegraphics[width=0.96\linewidth]{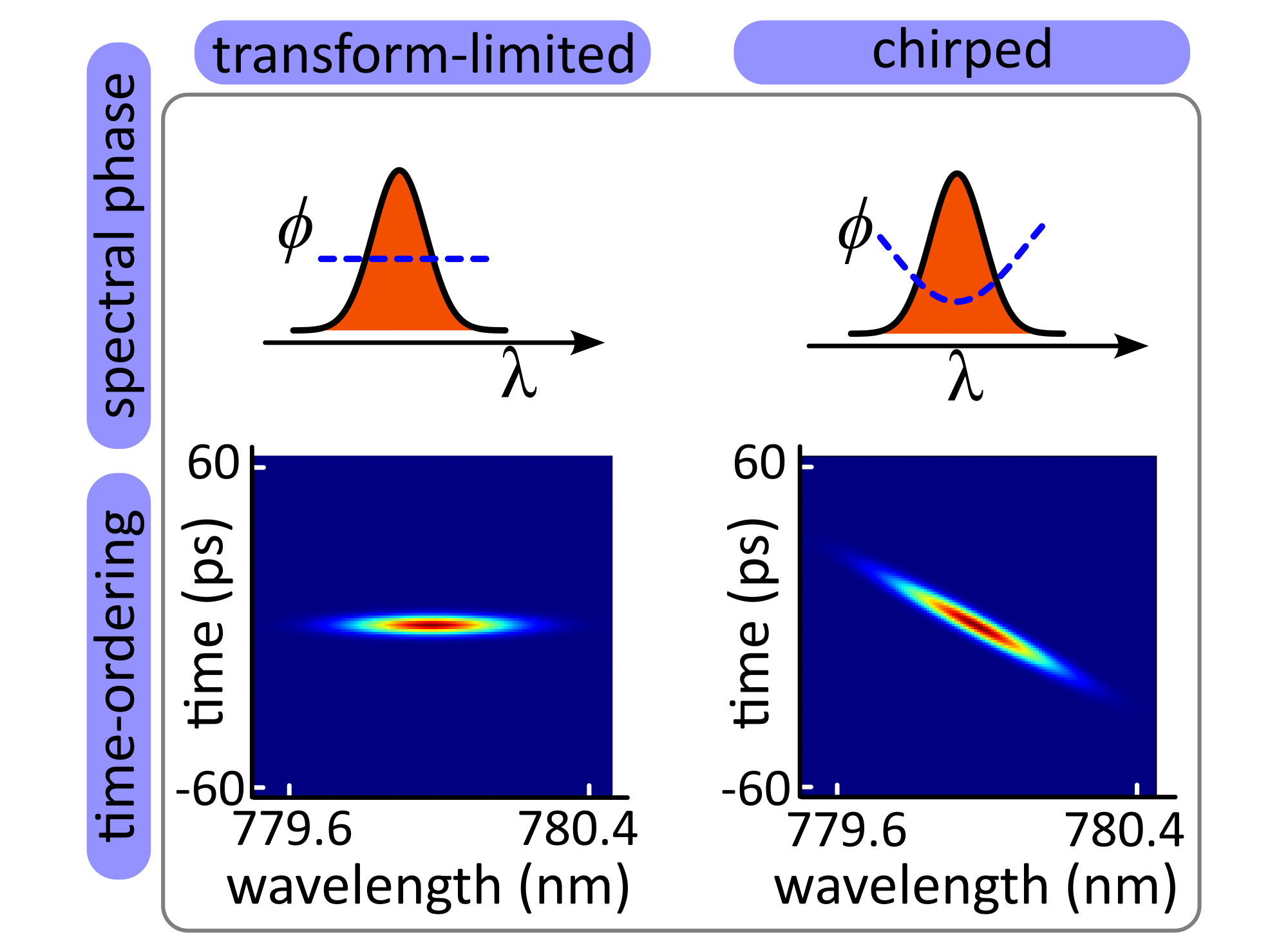}
    \caption{\textbf{Concept:} Representation of transform-limited and chirped laser pulses in spectral and temporal domains. The former has flat phase across the spectrum, while the latter carries a parabolic spectral phase. The corresponding wavelength-dependent arrival times are plotted in the bottom panel.}
 \label{fig_intro}
\end{figure}

Consider the frequency-dependent (spectral) phase of a broadband laser pulse, expressed as a Taylor series in angular frequency \cite{monmayrant2010newcomer}. The first-order component $\phi_1$ manifests as a time delay between pulse envelopes, the second-order component $\phi_2$ imparts a linear time delay between frequency components, dictating also their time ordering. 

The group delay $\tau_g(\omega)$ describes the derivative of the spectral phase $\phi(\omega)$ with respect to angular frequency:
\begin{equation}
   \tau_g(\omega) = \frac{d\phi(\omega)}{d\omega}.
\end{equation}
Group delay dispersion (GDD, also denoted as $\phi_2$) is the first derivative of group delay with respect to angular frequency, and also the second derivative of spectral phase with respect to angular frequency. 

The parameters $\phi_2$ and $\tau_g(\omega)$ are therefore related to each other as 
\begin{equation}
    \text{GDD} = \frac{d\tau_g}{d\omega} = \frac{d^2\phi(\omega)}{d\omega^2} = \phi_2.
\end{equation}

We can also define the so-called dispersion parameter $D_{1 \lambda}$ (in units of \si{\pico\second\per\nano\meter}), which is the wavelength-dependent-time-delay, related to GDD (in units of \si{\pico\second\squared}) through
\begin{equation}
   D_{1\lambda} = \frac{\partial T}{\partial \lambda} = -\frac{2 \pi c}{\lambda^2} \cdot \frac{\partial^2\phi}{\partial \omega^2} = -\frac{2 \pi c}{\lambda^2} \cdot \text{GDD}.
   \label{equ:gdd_time_lambda}
\end{equation}

The Figure \ref{fig_intro} summarizes this concept. In the absence of GDD, a broadband pulse maintains its temporal structure during propagation, with all spectral components travelling at identical group velocities, and hence have simultaneous arrival times. Under dispersive conditions, different wavelength components experience wavelength-dependent group velocities, leading to distinct arrival times. In other words, the temporal position of a wavelength directly encodes its wavelength information. As a result, by measuring the temporal position of spectral components, one can determine the GDD, at any power levels, as long as a photon wavepacket is detected, whose colour and arrival time can be determined accurately. For positive GDD, longer wavelengths arrive earlier than the blue ones, while for the negative chirp, it is vice versa. 

\section{Experimental scheme}

\begin{figure}[htb!]
	\centering
	\includegraphics[width=0.96\linewidth]{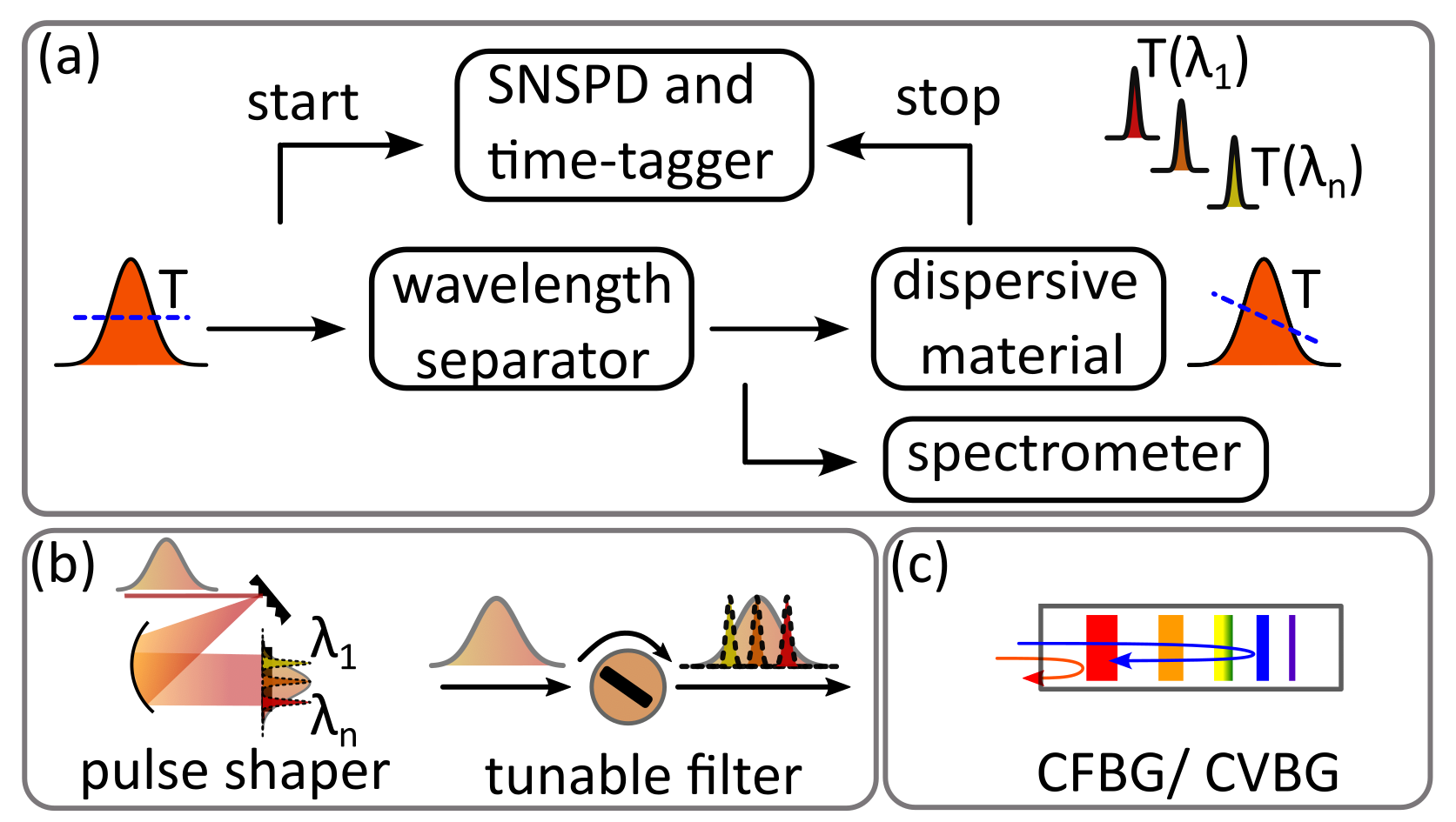}
    \caption{\textbf{Experimental scheme:}(a) A broadband laser source is directed through a wavelength separator, to tune the peak wavelength $\lambda_0$ and spectral width $\Delta \lambda$ of the spectral component measured using the spectrometer. The laser source triggers the reference \textit{start} signal and the chirped (i.e., time-delayed) spectral components trigger the \textit{stop} signals at the arrival time recorder consisting of single-photon detectors (here: superconducting nanowire single-photon detector, SNSPD) and time taggers, (b) chosen methods of wavelength separation, (c) dispersion methods: CF/VBG: chirped fiber/volume Bragg gratings.}
 \label{fig_exptscheme}
\end{figure}
\begin{figure*}[htb!]
	\centering
	\includegraphics[width=\linewidth]{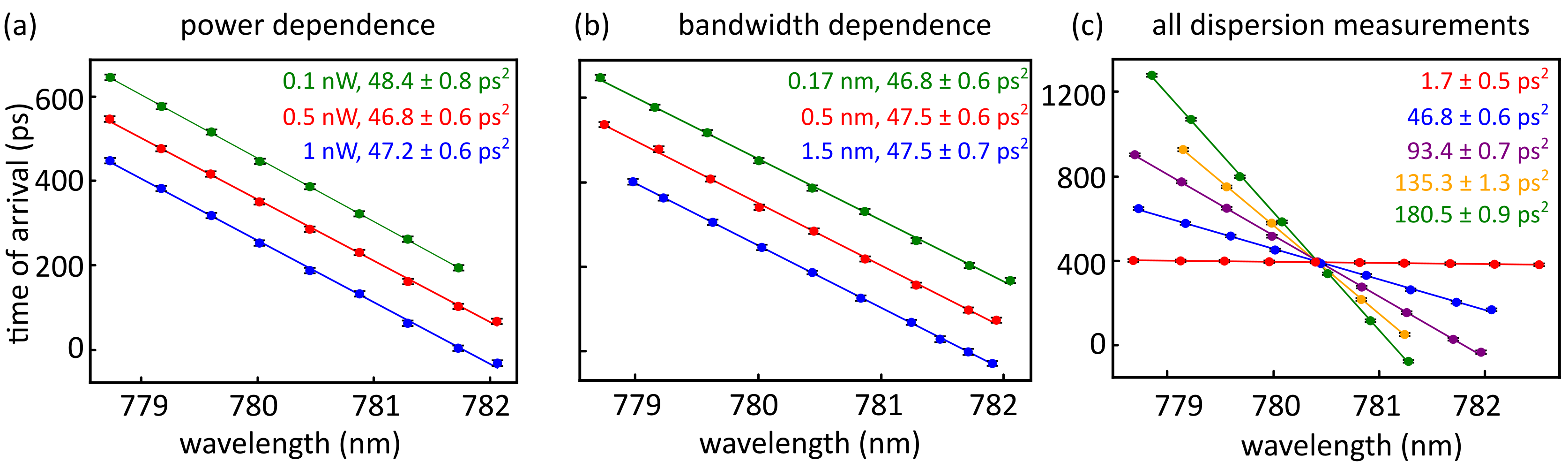}
    \caption{Results of the time-of-arrival measurement as a function of (a) laser power, (b) bandwidth. The measurement sets are vertically shifted for better visibility. (c) Results of all dispersion measurements. In all three cases, the filled circles denote data points and solid lines denote the linear fit, from which the GDD values are computed. }
 \label{fig_exptresults}
\end{figure*}
The experimental implementation of our technique is illustrated in Figure \ref{fig_exptscheme} (a). The setup consists of a pulsed laser source with a spectral bandwidth that matches that of the dispersion element. For the experiments reported here, we have employed a broadband laser source tuned to \SI{780}{\nano\meter} with transform-limited pulse duration \SI{170}{\femto\second}. As wavelength separator we couple the laser source to an automated, home-built 4$f$ pulse slicer (Figure \ref{fig_exptscheme} (b)) with a motorized slit, also enabling the control of spectral width. The design details are presented in Table \ref{tab1} and further details are in the Supplementary Information. Note that for wavelength selection, a tunable narrowband filter or grating filter can also be utilized (see results in Supplementary Information). 

\begin{table}[ht]
  \centering
  \begin{tabular}{l r} 
    \hline
    Item & Details \\
    \hline
    grating  & \SI{1800}{grooves/mm} \\
    focusing mirror & $ f=\SI{500}{\milli\meter}$  \\
    input beam size & \SI{1.7}{mm}  \\
    angle of incidence & \SI{38}{\degree}  \\
    \hline
    
    \textbf{resulting spectral resolution} & \SI{0.09}{nm}, for slit size \SI{130}{\micro\meter}  \\
    \hline
  \end{tabular}
  \caption{Design details of the tunable pulse shaper for spectral control. For further details see Ref. \cite{kappe2024chirped}}
  \label{tab1}
\end{table}
To imprint a well-defined GDD onto the laser pulse we use a chirped fiber and volume Bragg gratings (CF/VBG, see Figure \ref{fig_exptscheme} (c), and details of CFBG fabrication in Supplementary Information) with fixed dispersion values of \SI{45}{\pico\second\squared} at \SI{780}{nm} and their additive combinations . In chirped Bragg gratings, a linearly varying refractive index period is induced (along the propagation direction) via laser inscription \cite{remesh2023compact,glebov2014volume}, which provides time delays for wavelengths reflected from various positions. In principle, the dispersion can also be imparted using a programmable spatial light modulator. For spectral characterization, we employ a spectrometer with a charge-coupled device camera (Acton SP2750, Princeton Instruments, featuring an Andor iKon camera, with a spectral resolution \SI{0.02}{nm}). For temporal characterization, we use a high-time resolution superconducting nanowire single photon detector (SNSPD, Eos, Single Quantum, full width at half maximum timing jitter \SI{15}{ps}) associated with the time-tagging electronics (Time Tagger Ultra, Swabian Instruments, full width at half maximum timing jitter \SI{19}{ps}) to record time stamps of spectral components. 

The measurement routine is as follows: firstly, we direct the laser pulse along two measurement paths, a reference path to start the clock, setting time zero $T_0$, and the second branch with the 4$f$ pulse shaper that selects the spectral components. The spectral component propagating through the dispersive material is also sent to a spectrometer to determine its central wavelength and spectral width. The arrival times of spectral components that are time-delayed with respect to the reference clock ($T_{\lambda_{1...n}}$), are recorded concurrently using the SNSPD-time-tagger module. As the motor position in the pulse shaper is translated, wavelength components are scanned from smaller to larger values, and using the timing module, we record their corresponding arrival times. Effectively, this system helps correlate the spectral features with their corresponding temporal properties. For a positive chirp, the plot of wavelength vs arrival time would possess a negative slope, and vice versa. By fitting for a linear slope, we directly extract the dispersion parameter $D_{1\lambda}$, and based on Eq. \ref{equ:gdd_time_lambda}, compute the GDD value.

\section{Experimental Results}

We begin by investigating the technique at various laser power levels. We utilize the \SI{45}{\pico\second\squared} dispersion device and perform the wavelength-resolved time correlation measurement. The results are presented in Figure \ref{fig_exptresults} (a), where green, red and blue curves represent the average laser powers (at the input of the dispersion system) of \SI{0.1}{\nano\watt}, \SI{0.5}{\nano\watt}, and \SI{1}{\nano\watt} corresponding to \SI{1.25}{\atto\joule}, \SI{6.25}{\atto\joule}, and \SI{12.5}{\atto\joule} respectively. The individual measurement sets are vertically shifted for better visibility. The dots represent the data, and the solid line represents the linear fit. The $\lambda_{1...n}$ are computed from Gaussian fits on the shaped laser spectra, and the error bars are calculated from the central wavelength of the Gaussian distribution. Likewise, the presented arrival times $T_{\lambda_{1...n}}$ correspond to the means of Gaussian fits of four correlation measurements for each $\lambda_{n}$, and the error bars correspond to the standard deviation. The results show that all three measurements yield nearly the same slope, i.e., the same dispersion value.  The results indicate that the accuracy of the measurement procedure does not depend on the laser power employed. In principle, the measurement could be performed for even lower power levels, as long as the signal-to-noise ratio remains sufficient.

Next, we investigate the sensitivity of the technique to the spectral bandwidth. Using the controllable slit width in our pulse shaper, we choose three spectral widths \SI{0.1}{\nano\meter}, \SI{0.5}{\nano\meter}, and \SI{1}{\nano\meter} and perform the measurement, by appropriately controlling the slit width. The results are presented in Figure \ref{fig_exptresults} (b), with colors denoting the chosen spectral width. Yet again, we observe that measurements lead to nearly the same dispersion. To compute $\lambda_{1...n}$, and corresponding Gaussian fits are applied to the recorded spectra, and the time-correlation histograms and spectral profiles. When the latter deviate from Gaussian profiles, a suitable function must be identified to determine the median wavelength. For instance, in a 4$f$ pulse shaper, for certain slit sizes, the spectral response resembles a sigmoid-like function. In such cases, fitting with a Gaussian function is not ideal and must be adapted accordingly. 

Next, we investigate the sensitivity of the technique to measure various dispersion values. To this end, we set the input power to \SI{10}{nW}, and perform the measurements by combining CFBG and CVBG to obtain various linear combinations. The results are presented in Figure \ref{fig_exptresults} (c). The measured GDDs are \SI{46.8}{\pico\second\squared}, \SI{93.4}{\pico\second\squared}, \SI{135.3}{\pico\second\squared}, and \SI{180.5}{\pico\second\squared}, for the design values \SI{45}{\pico\second\squared}, \SI{90}{\pico\second\squared}, \SI{135}{\pico\second\squared}, and \SI{178}{\pico\second\squared} demonstrating a close match. From these measurements, it is clear that our experimental technique can measure large dispersion at the few-photon level. 

\begin{table*} [!hbt]
    \centering
    \begin{tabular}{ c }
    \includegraphics[width=\linewidth]{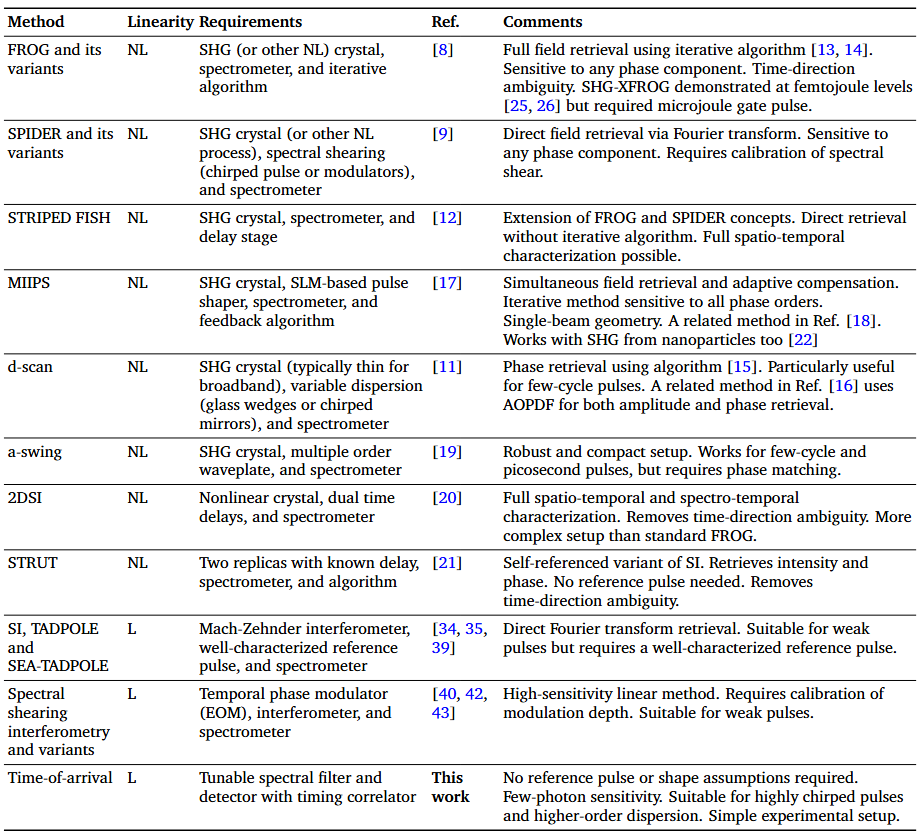}
\end{tabular}
\caption{\textbf{Comparison of ultrashort pulse characterization methods with their representative references}. NL: nonlinear; L: linear. Acronyms: FROG: Frequency-Resolved Optical Gating; SPIDER: Spectral Phase Interferometry for Direct Electric-field Reconstruction; STRIPED FISH: Spatially and Temporally Resolved Intensity and Phase Evaluation Device: Full Information from a Single Hologram; MIIPS: Multiphoton Intrapulse Interference Phase Scan; d-scan: dispersion scan; 2DSI: Two-Dimensional Spectral Interferometry; SI: Spectral Interferometry; STRUT: Spatially encoded arrangement for Temporal characterization Using a Translation stage; SEA TADPOLE: Spatially Encoded Arrangement for Temporal Analysis by Dispersing a Pair Of Light E-fields; SHG: second-harmonic generation; SLM: spatial light modulator; EOM: electro-optic modulator; AOM: acousto-optic modulator; AOPDF: acousto-optic programmable dispersive filter}
\label{tablechirpmethods}
\end{table*}

\section{Discussion and Conclusion} 
The accuracy of the time-of-arrival dispersion measurement technique, and in turn, the lowest dispersion that can be measured, may be affected by a few experimental factors. The primary error sources are the wavelength determination and arrival-time-recording systems. First, the precision of arrival time recording depends on the combined resolution of the SNSPD-time tagger system, which in our case is around \SI{24}{ps}. Here, the systematic uncertainties in the time-tagging electronics could introduce small measurement errors. 
Secondly, it is important to ensure a dispersion-free 4$f$ pulse shaper element for wavelength selection.
The time-of-arrival measurements without CF(V)BGs resulted in GDD = \SI{1.7}{\pico\second\squared} (see Figure \ref{fig_exptresults} (c)), which comes from the dispersion from the signal-delivering fiber to the detectors, non-zero dispersion of the 4$f$ pulse shaper, and experimental uncertainties or limitations. In our case, the dispersed spectral components travel around \SI{44}{m} distance through the signal delivery fiber to the detectors, contributing approximately \SI{2}{\pico\second\squared} dispersion. The remaining dispersion could be attributed to a few experimental limitations. The deviation of the 4$f$ pulse shaper from its perfect alignment results in non-zero dispersion (see extensive details in Ref. \cite{kappe2024chirped}). In our setup (see design details in Table \ref{tab1}), a deviation of \SI{1}{cm} from the ideal focusing condition only results in GDD = \SI{0.23}{\pico\second\squared}. As this value is smaller than the computed uncertainty values in our measurements, the technique is not sensitive in this range. We further investigated the sensitivity of the technique to the spectral range $\Delta \lambda$ and the number of wavelength sampling points within that range, and the results are presented in the Supplementary Information. Furthermore, real CF(V)BGs may exhibit slight variations from the designed GDD, due to variation in the apodization profiles, resulting from the manufacturing tolerances, leading to small departures from linear dispersion. For CFBGs, strain on the fiber, temperature, and material variations also play a role. Depending on the chosen slit size compared to the spectral resolution of the pulse shaper, the spectral shape may differ from Gaussian. Yet, as long as we can assign a $\lambda_{0}$ value and its corresponding arrival time, the technique may be adaptable to various dispersion measurement scenarios.

In Table \ref{tablechirpmethods}, we compare our time-of-arrival technique against the state-of-the-art dispersion characterization techniques. As we do not need any assumption about the pulse shape, nor require interference with a reference pulse, our time-of-arrival dispersion measurement technique is relatively simple and robust. Operating at few-photon levels, it is particularly well-suited for single-quantum emitter spectroscopy studies, high-dispersion characterization of on-chip photonic devices without the need for strong nonlinearity, and low-power optical communication systems.
The presented technique also bridges the gap between traditional ultrafast metrology and the quantum domain, establishing a pathway for dispersion characterization at power levels relevant to single-photon and few-photon optical technologies. Since it is a relatively fast technique that does not require high temporal resolution detectors, we can characterize arbitrarily large dispersion with a setup consisting of only a tunable filter and a detector coupled to a timing correlator.

Our wavelength-to-time mapping approach opens several promising avenues for future research. The technique's sensitivity at ultralow power levels makes it ideal for characterizing higher-order dispersion (third-order and beyond) in nonlinear optical systems where traditional high-power methods would induce unwanted nonlinear effects, particularly relevant for supercontinuum generation studies and soliton dynamics in optical fibers and waveguides \cite{Wetzel2018}. The method could be adapted for real-time spectroscopic monitoring of ultrafast processes \cite{Goda2013,Zhang2014} in biological systems and chemical reactions where sample photodamage limits available optical power. It could also be extended to time-resolved studies of transient dispersion phenomena in optically-excited materials, such as carrier-induced refractive index changes in semiconductors or photo-induced phase transitions, where conventional pump-probe techniques may be limited by sample damage thresholds. In the context of quantum photonics, wavelength-to-time mapping could facilitate the characterization of spectral diffusion and charge noise in solid-state quantum emitters by monitoring temporal jitter in photon arrival times as a function of wavelength. Furthermore, this approach may prove valuable for in-situ dispersion monitoring in long-haul fiber-optic communication networks, where real-time characterization without service interruption is critical.  

In summary, this work introduces and validates a linear, single-photon-sensitive technique for dispersion measurement that eliminates the need for nonlinear crystals, interferometers, or iterative phase-retrieval algorithms. By exploiting the wavelength-dependent temporal delay introduced by dispersive media and recording photon arrival times directly, the method can determine group delay dispersion at attojoule pulse energies. The excellent agreement between measured and nominal GDD values across a wide range of dispersion magnitudes and spectral bandwidths demonstrates both robustness and quantitative accuracy. Minor deviations can be attributed to fabrication tolerances in chirped Bragg gratings and to residual dispersion in the system. Importantly, the method is valid for both positive and negative second order dispersion, and for higher order dispersion factors. Our work opens the door for dispersion characterization in ultralow-energy quantum photonic systems, such as single-quantum-dot excitation and photon-pair generation, where nonlinear diagnostics are infeasible. Being general and easily reconfigurable, it can be adapted to other wavelength ranges and material systems. Combined with high-efficiency detectors and programmable spectral filters, it could enable in-situ pulse characterization in future integrated quantum photonic devices.

\begin{acknowledgments}
This research was funded in part by the Austrian Science Fund (FWF) projects with Grant DOIs 10.55776/COE1 (quantA) and 10.55776/FG5 (TM, MK, MT, IAA, RS, FK, YK, GW, and VR). Additionally, we acknowledge the infrastructure funding from FFG (HuSQI, grant number FO999896024). RGK and SN acknowledge financial support from the German Federal Ministry of Research, Technology, and Space through project 13N16028 (MHLASQU). For open access purposes, the authors have applied a CC BY public copyright license to any author-accepted manuscript version arising from this submission.
\end{acknowledgments}

\section*{Supplementary}

\subsection{CFBG details}
The CFBGs employed in this work are fabricated via femtosecond phase mask inscription technique \cite{thomas2012femtosecond} with \SI{800}{\nano\meter} wavelength, \SI{100}{\femto\second} pulse duration at \SI{200}{\hertz} repetition rate (Spectra-Physics Spitfire Ace). The fiber core refractive index modification is induced by multiphoton absorption of the incident laser irradiation. The non-periodic modulation of the refractive index modification is generated by a phase mask (i.e., a transmission grating, manufactured in-house via electron-beam lithography at the University of Jena), which defines the resulting distribution of the grating period along the fiber axis. 

\subsection{Pulse shaper design}
A 4$f$ shaper relies on spatial mapping of frequency components and their subsequent control using a mask \cite{monmayrant2010newcomer,kappe2024chirped}. The popular implementation is a folded version, combining a diffraction grating and a focusing mirror to angularly disperse and and collect the frequency components of the input and the output pulses, and to map individual spectral components to a spatial coordinate (Fourier plane) respectively. Assuming a linear dispersion of the diffraction grating, the position of a single frequency component $\omega_k$ at the Fourier plane is given by $X_k=\alpha \omega_k$, where 
\begin{equation}
    \alpha = \frac{\lambda_0^2 f}{2 \pi c d \cos \theta_d}
\end{equation}
is defined by the parameters of the 4$f$ shaper: $d$ the grating period, $\lambda_0$ the central wavelength, $\theta_d$ the diffraction angle of the grating, and $c$ the speed of light. The spot size of a single frequency component at the Fourier plane is given by
\begin{equation}
\Delta x_0 =2 \ln (2) \frac{\cos \theta_i}{\cos \theta_d} \frac{f \lambda_0}{\pi \Delta x_{\mathrm{in}}},
\end{equation},
where $\Delta x_{\mathrm{in}}$ is the spot size at the grating plane and $\theta_i$ is the angle of incidence. Using these two equations, the frequency resolution at the Fourier plane is obtained as 
\begin{equation}
    \delta \omega=\Delta x_0 / \alpha.
    \label{equ:frequ_resolution_slm}
\end{equation}
The optimal resolution is determined by the input beam size, the period of the grating, and the ratio $\cos{\theta_{i}}/\cos{\theta_{d}}$. To control the Fourier plane spot size, one may also tune the angle of incidence at the grating. To improve the resolution, one can increase the focal length, allowing finer spectral features to be resolved, increase the beam size to reduce the diffraction-induced broadening, and improve spectral selectivity, or use a higher groove density grating, which increases angular dispersion, thereby spreading the spectral components further apart. 

\subsection{Wavelength selecting method}
\begin{figure}[htb!]
	\centering
	\includegraphics[width=\linewidth]{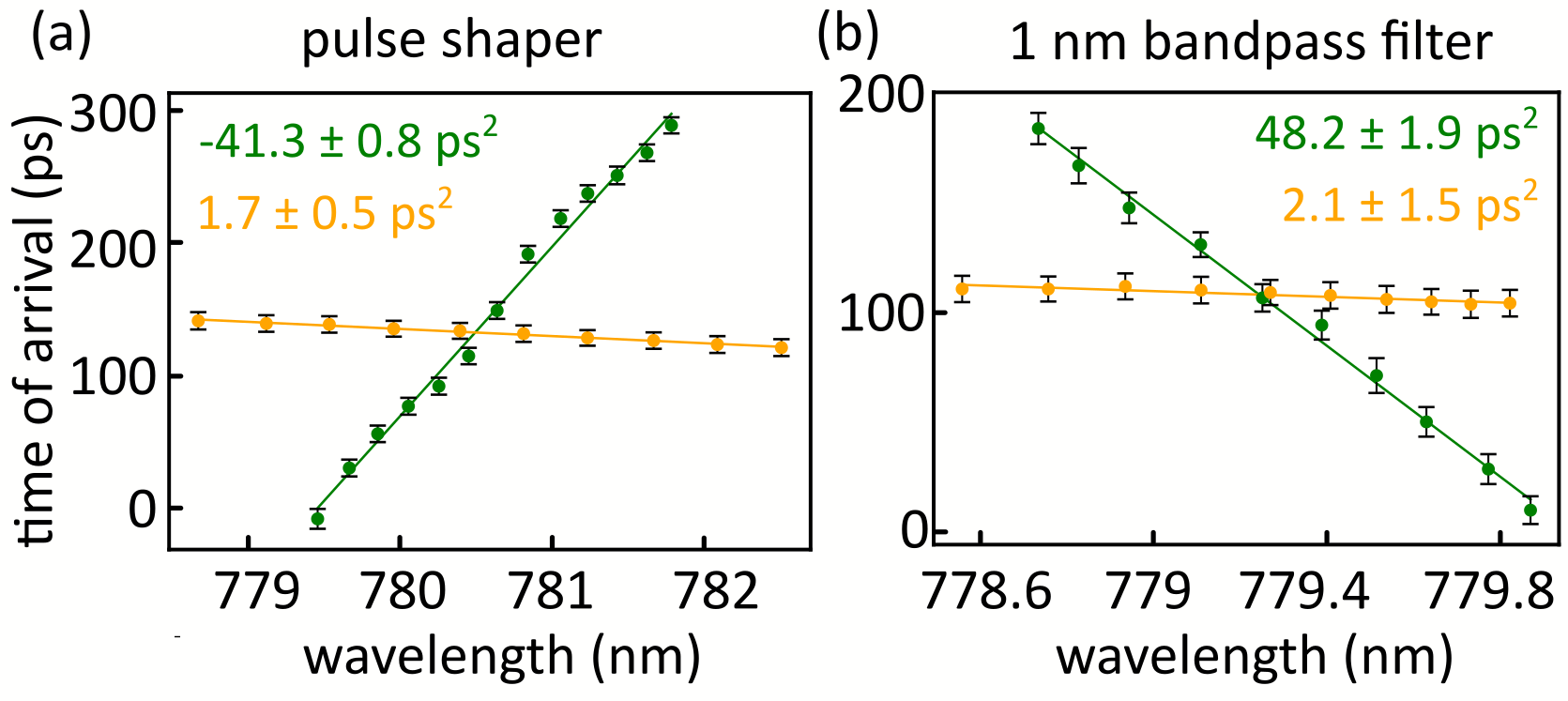}
    \caption{Results of time-of-arrival dispersion measurement with (a) pulse shaper, and (b) tunable bandpass filter as wavelength selector device. }
 \label{fig_bandpass}
\end{figure}
We validated the adaptability of time-of-arrival method with two different wavelength selecting techniques: a pulse shaper and a tunable bandpass filter. The results are presented in Figure \ref{fig_bandpass} (a) and (b), which establishes that the presented method does not necessarily need a 4$f$ pulse shaping apparatus. 

\subsection{Dependence on spectral bandwidth and wavelength samples at very low dispersion}
\begin{figure}[htb!]
	\centering
	\includegraphics[width=\linewidth]{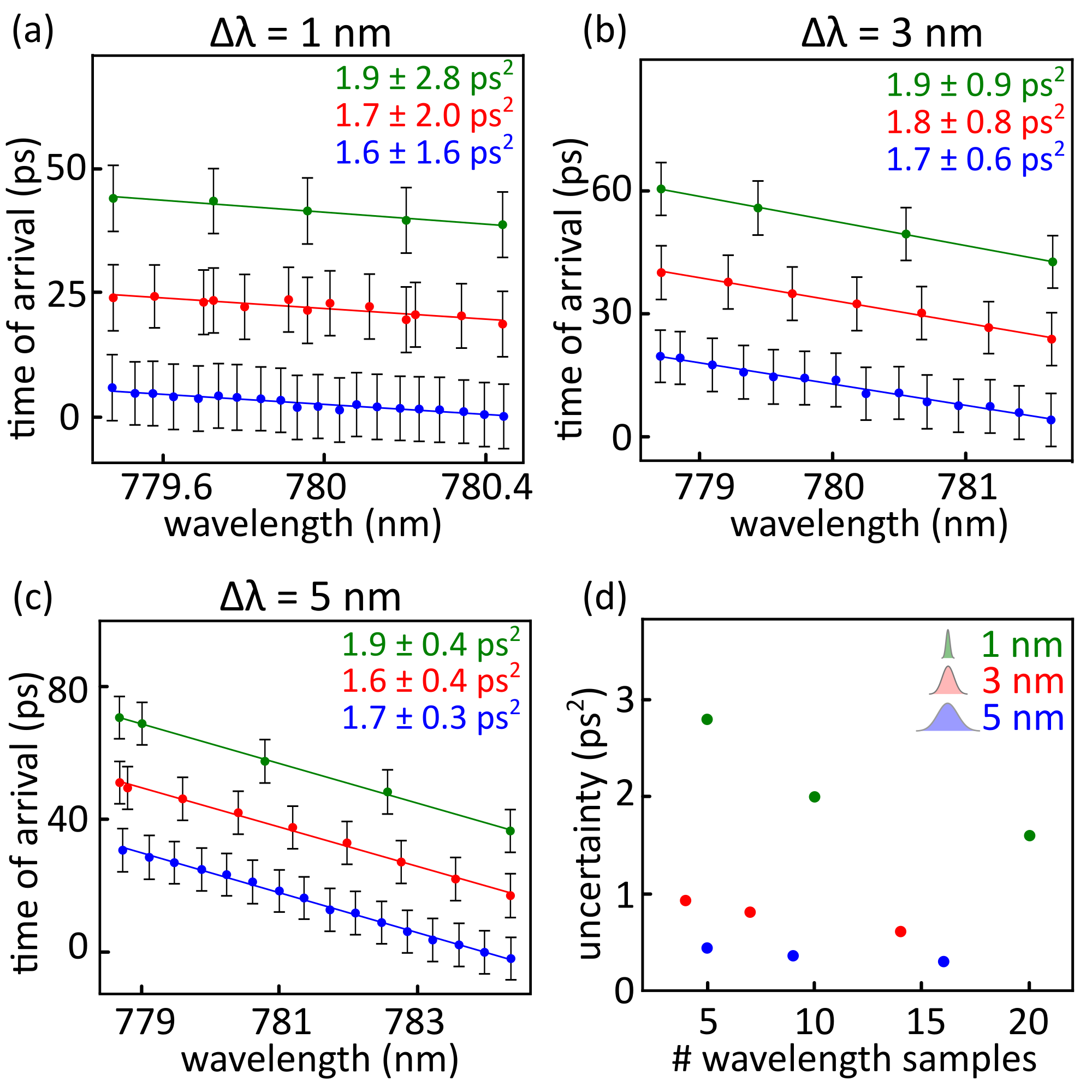}
    \caption{(a)-(c) Results of time-of-arrival dispersion measurement as a function of spectral bandwidth and number of wavelength samples (d) Summary of (a)-(c), i.e., computed uncertainty as a function of wavelength samples for various spectral bandwidths. }
 \label{fig_limits}
\end{figure}

We investigated the sensitivity of the time-of-arrival method to measure very low dispersion, as a function of laser spectral bandwidths and the number of wavelength samples required in the experimental routine. The results are presented in Figure \ref{fig_limits} (a)-(d). We observe that for a given measurement set with a fixed number of wavelength samples, the computed uncertainty in the GDD measurement is inversely related to the spectral bandwidth: the smaller the bandwidth, the larger the uncertainty in the measurement. In general, smaller uncertainty is obtained for larger number of wavelength samples. 

\section{References}
\bibliography{bib}%

@article{warren1993coherent,
  title = {Coherent Control of Quantum Dynamics: The Dream Is Alive},
  volume = {259},
  ISSN = {1095-9203},
  url = {http://dx.doi.org/10.1126/science.259.5101.1581},
  DOI = {10.1126/science.259.5101.1581},
  number = {5101},
  journal = {Science},
  publisher = {American Association for the Advancement of Science (AAAS)},
  author = {Warren,  Warren S. and Rabitz,  Herschel and Dahleh,  Mohammed},
  year = {1993},
  pages = {1581–1589}
}

@article{chatel_competition_2003,
  title = {Competition between sequential and direct paths in a two-photon transition},
  volume = {68},
  ISSN = {1094-1622},
  url = {http://dx.doi.org/10.1103/PhysRevA.68.041402},
  DOI = {10.1103/PhysRevA.68.041402},
  number = {4},
  journal = {Physical Review A},
  publisher = {American Physical Society (APS)},
  author = {Chatel,  Beatrice and Degert,  Jerome and Stock,  Sabine and Girard,  Bertrand},
  year = {2003}, 
}

@article{remesh2023compact,
    author = {Remesh, Vikas and Krämer, Ria G. and Schwarz, René and Kappe, Florian and Karli, Yusuf and Siems, Malte Per and Bracht, Thomas K. and Covre da Silva, Saimon Filipe and Rastelli, Armando and Reiter, Doris E. and Richter, Daniel and Nolte, Stefan and Weihs, Gregor},
    title = "{Compact chirped fiber Bragg gratings for single-photon generation from quantum dots}",
    journal = {APL Photonics},
    volume = {8},
    number = {10},
    pages = {101301},
    year = {2023},
    month = {10},
    issn = {2378-0967},
    doi = {10.1063/5.0164222},
    url = {https://doi.org/10.1063/5.0164222},
}

@article{kappe2024chirped,
  title = {Chirped Pulses Meet Quantum Dots: Innovations,  Challenges,  and Future Perspectives},
  volume = {8},
  ISSN = {2511-9044},
  url = {http://dx.doi.org/10.1002/qute.202300352},
  DOI = {10.1002/qute.202300352},
  number = {2},
  journal = {Advanced Quantum Technologies},
  publisher = {Wiley},
  author = {Kappe,  Florian and Karli,  Yusuf and Wilbur,  Grant and Kr\"{a}mer,  Ria G. and Ghosh,  Sayan and Schwarz,  René and Kaiser,  Moritz and Bracht,  Thomas K. and Reiter,  Doris E. and Nolte,  Stefan and Hall,  Kimberley C. and Weihs,  Gregor and Remesh,  Vikas},
  year = {2024},
  month = jan 
}

@article{ramachandran_experimental_2021,
author = {A. Ramachandran and J. Fraser-Leach and S. O'Neal and D. G. Deppe and K. C. Hall},
journal = {Opt. Express},
number = {25},
pages = {41766--41775},
publisher = {Optica Publishing Group},
title = {Experimental quantification of the robustness of adiabatic rapid passage for quantum state inversion in semiconductor quantum dots},
volume = {29},
month = {Dec},
year = {2021},
url = {https://opg.optica.org/oe/abstract.cfm?URI=oe-29-25-41766},
doi = {10.1364/OE.435109},
}

@article{kappe2022collective,
doi = {10.1088/2633-4356/acd7c1},
url = {https://dx.doi.org/10.1088/2633-4356/acd7c1},
year = {2023},
month = {jun},
publisher = {IOP Publishing},
volume = {3},
number = {2},
pages = {025006},
author = {Florian Kappe and Yusuf Karli and Thomas K Bracht and Saimon Filipe Covre da Silva and Tim Seidelmann and Vollrath Martin Axt and Armando Rastelli and Gregor Weihs and Doris E Reiter and Vikas Remesh},
title = {Collective excitation of spatio-spectrally distinct quantum dots enabled by chirped pulses},
journal = {Mater. Quantum Technol.}
}

@article{kurzyna2022variable,
  title={Variable electro-optic shearing interferometry for ultrafast single-photon-level pulse characterization},
  author={Kurzyna, Stanis{\l}aw and Jastrz{\k{e}}bski, Marcin and Fabre, Nicolas and Wasilewski, Wojciech and Lipka, Micha{\l} and Parniak, Micha{\l}},
  journal={Optics Express},
  volume={30},
  number={22},
  pages={39826--39839},
  year={2022},
doi = {10.1364/OE.471108},
url = {https://doi.org/10.1364/OE.471108},
  publisher={Optica Publishing Group}
}

@article{lepetit1995linear,
  title={Linear techniques of phase measurement by femtosecond spectral interferometry for applications in spectroscopy},
  author={Lepetit, L and Ch{\'e}riaux, G and Joffre, M},
  journal={JOSA B},
  volume={12},
  number={12},
  pages={2467--2474},
  year={1995},
doi = {10.1364/JOSAB.12.002467},
url = {https://doi.org/10.1364/JOSAB.12.002467},
  publisher={Optica Publishing Group}
}

@article{trebino_measuring_1997,
	title = {Measuring ultrashort laser pulses in the time-frequency domain using frequency-resolved optical gating},
	volume = {68},
	issn = {0034-6748, 1089-7623},
	url = {http://aip.scitation.org/doi/10.1063/1.1148286},
	doi = {10.1063/1.1148286},
	language = {},
	number = {9},
	urldate = {2022-08-22},
	journal = {Review of Scientific Instruments},
	author = {Trebino, Rick and DeLong, Kenneth W. and Fittinghoff, David N. and Sweetser, John N. and Krumbügel, Marco A. and Richman, Bruce A. and Kane, Daniel J.},
	month = sep,
	year = {1997},
	pages = {3277--3295},
}

@article{chilla1991direct,
  title={Direct determination of the amplitude and the phase of femtosecond light pulses},
  author={Chilla, Juan LA and Martinez, Oscar E},
  journal={Optics letters},
  volume={16},
  number={1},
  pages={39--41},
  year={1991},
doi = {10.1364/OL.16.000039},
url = {https://doi.org/10.1364/OL.16.000039},
  publisher={Optical Society of America}
}

@article{iaconis1998spectral,
  title = {Spectral phase interferometry for direct electric-field reconstruction of ultrashort optical pulses},
  volume = {23},
  ISSN = {1539-4794},
  url = {http://dx.doi.org/10.1364/OL.23.000792},
  DOI = {10.1364/ol.23.000792},
  number = {10},
  journal = {Optics Letters},
  publisher = {Optica Publishing Group},
  author = {Iaconis,  C. and Walmsley,  I. A.},
  year = {1998},
  month = may,
  pages = {792}
}

@article{stibenz2005interferometric,
  title = {Interferometric frequency-resolved optical gating},
  volume = {13},
  ISSN = {1094-4087},
  url = {http://dx.doi.org/10.1364/OPEX.13.002617},
  DOI = {10.1364/opex.13.002617},
  number = {7},
  journal = {Optics Express},
  publisher = {Optica Publishing Group},
  author = {Stibenz,  Gero and Steinmeyer,  G�nter},
  year = {2005},
  pages = {2617}
}

@article{lozovoy2004multiphoton,
  title = {Multiphoton intrapulse interference-IV-Ultrashort laser pulse spectral phase characterization and compensation},
  volume = {29},
  ISSN = {1539-4794},
  url = {http://dx.doi.org/10.1364/OL.29.000775},
  DOI = {10.1364/ol.29.000775},
  number = {7},
  journal = {Optics Letters},
  publisher = {Optica Publishing Group},
  author = {Lozovoy,  Vadim V. and Pastirk,  Igor and Dantus,  Marcos},
  year = {2004},
  month = apr,
  pages = {775}
}

@article{miranda2012characterization,
  title = {Characterization of broadband few-cycle laser pulses with the d-scan technique},
  volume = {20},
  ISSN = {1094-4087},
  url = {http://dx.doi.org/10.1364/OE.20.018732},
  DOI = {10.1364/oe.20.018732},
  number = {17},
  journal = {Optics Express},
  publisher = {Optica Publishing Group},
  author = {Miranda,  Miguel and Arnold,  Cord L. and Fordell,  Thomas and Silva,  Francisco and Alonso,  Benjamín and Weigand,  Rosa and L’Huillier,  Anne and Crespo,  Helder},
  year = {2012},
  month = aug,
  pages = {18732}
}

@article{remesh_coherent_2019,
	title = {Coherent {Multiphoton} {Control} of {Gallium} {Phosphide} {Nanodisk} {Resonances}},
	volume = {6},
	issn = {2330-4022, 2330-4022},
	url = {https://pubs.acs.org/doi/10.1021/acsphotonics.9b00780},
	doi = {10.1021/acsphotonics.9b00780},
	language = {},
	number = {10},
	urldate = {2022-08-22},
	journal = {ACS Photonics},
	author = {Remesh, Vikas and Grinblat, Gustavo and Li, Yi and Maier, Stefan A. and van Hulst, Niek F.},
	month = oct,
	year = {2019},
	pages = {2487--2491},
}

@article{backus_high_1998,
	title = {High power ultrafast lasers},
	volume = {69},
	issn = {0034-6748, 1089-7623},
	url = {http://aip.scitation.org/doi/10.1063/1.1148795},
	doi = {10.1063/1.1148795},
	language = {},
	number = {3},
	urldate = {2022-08-09},
	journal = {Review of Scientific Instruments},
	author = {Backus, Sterling and Durfee, Charles G. and Murnane, Margaret M. and Kapteyn, Henry C.},
	month = mar,
	year = {1998},
	pages = {1207--1223},
}

@article{ranka1997autocorrelation,
  title={Autocorrelation measurement of 6-fs pulses based on the two-photon-induced photocurrent in a GaAsP photodiode},
  author={Ranka, Jinendra K and Gaeta, Alexander L and Baltuska, Andrius and Pshenichnikov, Maxim S and Wiersma, Douwe A},
  journal={Opt. Lett.},
  volume={22},
  number={17},
  pages={1344--1346},
  year={1997},
url          = {https://doi.org/10.1364/OL.22.001344},
	doi      = {10.1364/OL.22.001344},
  publisher={Optica Publishing Group}
}

@article{chong2014autocorrelation,
  title={Autocorrelation measurement of femtosecond laser pulses based on two-photon absorption in GaP photodiode},
  author={Chong, EZ and Watson, TF and Festy, F},
  journal={Appl. Phys.},
  volume={105},
  number={6},
  pages={062111},
  year={2014},
url          = {https://doi.org/10.1063/1.4893423},
	doi      = {10.1063/1.4893423},
  publisher={AIP Publishing LLC}
}

@article{kleimeier2010autocorrelation,
  title={Autocorrelation and phase retrieval in the UV using two-photon absorption in diamond pin photodiodes},
  author={Kleimeier, Nils Fabian and Haarlammert, Thorben and Witte, Henrik and Sch{\"u}hle, Udo and Hochedez, Jean-Francois and BenMoussa, Ali and Zacharias, Helmut},
  journal={Optics express},
  volume={18},
  number={7},
  pages={6945--6956},
  year={2010},
url          = {https://doi.org/10.1364/OE.18.006945},
	doi          = {10.1364/OE.18.006945},
  publisher={Optica Publishing Group}
}

@article{geindre2001single,
  title={Single-shot spectral interferometry with chirped pulses},
  author={Geindre, J-P and Audebert, P and Rebibo, S and Gauthier, J-C},
  journal={Opt. Lett. },
  volume={26},
  number={20},
  pages={1612--1614},
  year={2001},
url          = {https://doi.org/10.1364/OL.26.001612},
	doi          = {10.1364/OL.26.001612},
  publisher={Optica Publishing Group}
}

@article{sainz1994real,
  title={Real time interferometric measurements of dispersion curves},
  author={Sainz, C and Jourdian, P and Escalona, R and Calatroni, J},
  journal={Optics Communications},
  volume={110},
  number={3-4},
  pages={381--390},
  year={1994},
url          = {https://doi.org/10.1016/0030-4018(94)90442-1},
	doi          = {10.1016/0030-4018(94)90442-1},
  publisher={Elsevier}
}

@article{kovacs2005dispersion,
  title={Dispersion control of a pulse stretcher--compressor system with two-dimensional spectral interferometry},
  author={Kov{\'a}cs, AP and Osvay, K and Kurdi, G and G{\"o}rbe, M and Klebniczki, J and Bor, Zs},
  journal={Appl. Phys. B},
  volume={80},
  pages={165--170},
  year={2005},
url          = { https://doi.org/10.1007/s00340-004-1706-0},
	doi          = {10.1007/s00340-004-1706-0},
  publisher={Springer}
}

@article{fan_measurement_2013,
	title = {Measurement of the chirp characteristics of linearly chirped pulses by a frequency domain interference method},
	volume = {21},
	issn = {1094-4087},
	url = {https://opg.optica.org/oe/abstract.cfm?uri=oe-21-11-13062},
	doi = {10.1364/OE.21.013062},
	language = {},
	number = {11},
	urldate = {2022-08-10},
	journal = {Optics Express},
	author = {Fan, Wei and Zhu, Bin and Wu, Yinzhong and Qian, Feng and Shui, Min and Du, Sai and Zhang, Bo and Wu, Yuchi and Xin, Jianting and Zhao, Zongqing and Cao, Leifeng and Wang, Yuxiao and Gu, Yuqiu},
	month = jun,
	year = {2013},
	pages = {13062},
}

@article{dorrer2008linear,
  title = {Linear self-referencing techniques for short-optical-pulse characterization [Invited]},
  volume = {25},
  ISSN = {1520-8540},
  url = {http://dx.doi.org/10.1364/JOSAB.25.0000A1},
  DOI = {10.1364/josab.25.0000a1},
  number = {6},
  journal = {Journal of the Optical Society of America B},
  publisher = {Optica Publishing Group},
  author = {Dorrer,  C. and Kang,  I.},
  year = {2008},
  month = mar,
  pages = {A1}
}

@article{glebov2014volume,
  title = {Volume-chirped Bragg gratings: monolithic components for stretching and compression of ultrashort laser pulses},
  volume = {53},
  ISSN = {0091-3286},
  url = {http://dx.doi.org/10.1117/1.OE.53.5.051514},
  DOI = {10.1117/1.oe.53.5.051514},
  number = {5},
  journal = {Optical Engineering},
  publisher = {SPIE-Intl Soc Optical Eng},
  author = {Glebov,  Leonid and Smirnov,  Vadim and Rotari,  Eugeniu and Cohanoschi,  Ion and Glebova,  Larissa and Smolski,  Oleg and Lumeau,  Julien and Lantigua,  Christopher and Glebov,  Alexei},
  year = {2014},
  month = feb,
  pages = {051514}
}

@article{Zhang2003,
  title = {Measurement of the intensity and phase of attojoule femtosecond light pulses using Optical-Parametric-Amplification Cross-Correlation Frequency-Resolved Optical Gating},
  volume = {11},
  ISSN = {1094-4087},
  url = {http://dx.doi.org/10.1364/OE.11.000601},
  DOI = {10.1364/oe.11.000601},
  number = {6},
  journal = {Optics Express},
  publisher = {Optica Publishing Group},
  author = {Zhang,  Jing-yuan and Shreenath,  Aparna and Kimmel,  Mark and Zeek,  Erik and Trebino,  Rick and Link,  Stephan},
  year = {2003},
  month = mar,
  pages = {601}
}

@article{Zacharias2025,
  title = {Energy-Efficient Ultrashort-Pulse Characterization Using Nanophotonic Parametric Amplification},
  volume = {12},
  ISSN = {2330-4022},
  url = {http://dx.doi.org/10.1021/acsphotonics.4c02620},
  DOI = {10.1021/acsphotonics.4c02620},
  number = {3},
  journal = {ACS Photonics},
  publisher = {American Chemical Society (ACS)},
  author = {Zacharias,  Thomas and Gray,  Robert and Sekine,  Ryoto and Williams,  James and Zhou,  Selina and Marandi,  Alireza},
  year = {2025},
  month = mar,
  pages = {1316–1320}
}

@article{Itatani2002,
  title = {Attosecond Streak Camera},
  volume = {88},
  ISSN = {1079-7114},
  url = {http://dx.doi.org/10.1103/PhysRevLett.88.173903},
  DOI = {10.1103/physrevlett.88.173903},
  number = {17},
  journal = {Physical Review Letters},
  publisher = {American Physical Society (APS)},
  author = {Itatani,  J. and Quéré,  F. and Yudin,  G. L. and Ivanov,  M. Yu. and Krausz,  F. and Corkum,  P. B.},
  year = {2002},
  month = apr 
}

@article{Park2018,
  title = {Direct sampling of a light wave in air},
  volume = {5},
  ISSN = {2334-2536},
  url = {http://dx.doi.org/10.1364/OPTICA.5.000402},
  DOI = {10.1364/optica.5.000402},
  number = {4},
  journal = {Optica},
  publisher = {Optica Publishing Group},
  author = {Park,  Seung Beom and Kim,  Kyungseung and Cho,  Wosik and Hwang,  Sung In and Ivanov,  Igor and Nam,  Chang Hee and Kim,  Kyung Taec},
  year = {2018},
  month = apr,
  pages = {402}
}

@article{Kaldewey2017,
  title = {Coherent and robust high-fidelity generation of a biexciton in a quantum dot by rapid adiabatic passage},
  volume = {95},
  ISSN = {2469-9969},
  url = {http://dx.doi.org/10.1103/PhysRevB.95.161302},
  DOI = {10.1103/physrevb.95.161302},
  number = {16},
  journal = {Physical Review B},
  publisher = {American Physical Society (APS)},
  author = {Kaldewey,  Timo and L\"{u}ker,  Sebastian and Kuhlmann,  Andreas V. and Valentin,  Sascha R. and Ludwig,  Arne and Wieck,  Andreas D. and Reiter,  Doris E. and Kuhn,  Tilmann and Warburton,  Richard J.},
  year = {2017},
  month = apr 
}

@article{Kaldewey2018,
  title = {Far-field nanoscopy on a semiconductor quantum dot via a rapid-adiabatic-passage-based switch},
  volume = {12},
  ISSN = {1749-4893},
  url = {http://dx.doi.org/10.1038/s41566-017-0079-y},
  DOI = {10.1038/s41566-017-0079-y},
  number = {2},
  journal = {Nature Photonics},
  publisher = {Springer Science and Business Media LLC},
  author = {Kaldewey,  Timo and Kuhlmann,  Andreas V. and Valentin,  Sascha R. and Ludwig,  Arne and Wieck,  Andreas D. and Warburton,  Richard J.},
  year = {2018},
  month = jan,
  pages = {68–72}
}

@article{Wetzel2018,
  title = {Customizing supercontinuum generation via on-chip adaptive temporal pulse-splitting},
  volume = {9},
  ISSN = {2041-1723},
  url = {http://dx.doi.org/10.1038/s41467-018-07141-w},
  DOI = {10.1038/s41467-018-07141-w},
  number = {1},
  journal = {Nature Communications},
  publisher = {Springer Science and Business Media LLC},
  author = {Wetzel,  Benjamin and Kues,  Michael and Roztocki,  Piotr and Reimer,  Christian and Godin,  Pierre-Luc and Rowley,  Maxwell and Little,  Brent E. and Chu,  Sai T. and Viktorov,  Evgeny A. and Moss,  David J. and Pasquazi,  Alessia and Peccianti,  Marco and Morandotti,  Roberto},
  year = {2018},
  month = nov 
}

@article{Gabolde2006,
  title = {Single-shot measurement of the full spatio-temporal field of ultrashort pulses with multi-spectral digital holography},
  volume = {14},
  ISSN = {1094-4087},
  url = {http://dx.doi.org/10.1364/OE.14.011460},
  DOI = {10.1364/oe.14.011460},
  number = {23},
  journal = {Optics Express},
  publisher = {Optica Publishing Group},
  author = {Gabolde,  Pablo and Trebino,  Rick},
  year = {2006},
  pages = {11460}
}

@article{Birge2006,
  title = {Two-dimensional spectral shearing interferometry for few-cycle pulse characterization},
  volume = {31},
  ISSN = {1539-4794},
  url = {http://dx.doi.org/10.1364/OL.31.002063},
  DOI = {10.1364/ol.31.002063},
  number = {13},
  journal = {Optics Letters},
  publisher = {Optica Publishing Group},
  author = {Birge,  Jonathan R. and Ell,  Richard and K\"{a}rtner,  Franz X.},
  year = {2006},
  month = jul,
  pages = {2063}
}

@article{Geib2019,
  title = {Common pulse retrieval algorithm: a fast and universal method to retrieve ultrashort pulses},
  volume = {6},
  ISSN = {2334-2536},
  url = {http://dx.doi.org/10.1364/OPTICA.6.000495},
  DOI = {10.1364/optica.6.000495},
  number = {4},
  journal = {Optica},
  publisher = {Optica Publishing Group},
  author = {Geib,  Nils C. and Zilk,  Matthias and Pertsch,  Thomas and Eilenberger,  Falk},
  year = {2019},
  month = apr,
  pages = {495}
}

@article{thomas2012femtosecond,
  title={Femtosecond pulse written fiber gratings: a new avenue to integrated fiber technology},
  author={Thomas, Jens and Voigtlaender, Christian and Becker, Ria G and Richter, Daniel and Tuennermann, Andreas and Nolte, Stefan},
  journal={Laser Photonics Rev.},
  volume={6},
  number={6},
url  = {https://doi.org/10.1002/lpor.201100033},
	doi          = {10.1002/lpor.201100033}, 
  pages={709--723},
  year={2012},
  publisher={Wiley Online Library}
}

@article{monmayrant2010newcomer,
	author       = {Monmayrant, Antoine and Weber, S{\'{e}}bastien and Chatel, B{\'{e}}atrice},
	issn         = {09534075},
	journal      = {J. Phys. B At. Mol. Opt. Phys.},
	number       = {10},
	pages        = {103001},
	publisher    = {IOP Publishing},
	title        = {{A newcomer's guide to ultrashort pulse shaping and characterization}},
	volume       = {43},
	year         = {2010},
	doi          = {10.1088/0953-4075/43/10/103001},
    url          = {https://dx.doi.org/10.1088/0953-4075/43/10/103001},
}

@article{Goda2013,
  title = {Dispersive Fourier transformation for fast continuous single-shot measurements},
  volume = {7},
  ISSN = {1749-4893},
  url = {http://dx.doi.org/10.1038/nphoton.2012.359},
  DOI = {10.1038/nphoton.2012.359},
  number = {2},
  journal = {Nature Photonics},
  publisher = {Springer Science and Business Media LLC},
  author = {Goda,  K. and Jalali,  B.},
  year = {2013},
  month = jan,
  pages = {102–112}
}

@article{Zhang2014,
  title = {Ultrafast and versatile spectroscopy by temporal Fourier transform},
  volume = {4},
  ISSN = {2045-2322},
  url = {http://dx.doi.org/10.1038/srep05351},
  DOI = {10.1038/srep05351},
  number = {1},
  journal = {Scientific Reports},
  publisher = {Springer Science and Business Media LLC},
  author = {Zhang,  Chi and Wei,  Xiaoming and Marhic,  Michel E. and Wong,  Kenneth K. Y.},
  year = {2014},
  month = jun 
}

@article{DeLong1994,
  title = {Pulse retrieval in frequency-resolved optical gating based on the method of generalized projections},
  volume = {19},
  ISSN = {1539-4794},
  url = {http://dx.doi.org/10.1364/OL.19.002152},
  DOI = {10.1364/ol.19.002152},
  number = {24},
  journal = {Optics Letters},
  publisher = {Optica Publishing Group},
  author = {DeLong,  Kenneth W. and Kohler,  Bern and Wilson,  Kent and Fittinghoff,  David N. and Trebino,  Rick},
  year = {1994},
  month = dec,
  pages = {2152}
}

@article{Miranda2016,
  title = {Fast iterative retrieval algorithm for ultrashort pulse characterization using dispersion scans},
  volume = {34},
  ISSN = {1520-8540},
  url = {http://dx.doi.org/10.1364/JOSAB.34.000190},
  DOI = {10.1364/josab.34.000190},
  number = {1},
  journal = {Journal of the Optical Society of America B},
  publisher = {Optica Publishing Group},
  author = {Miranda,  Miguel and Penedones,  João and Guo,  Chen and Harth,  Anne and Louisy,  Maïté and Neoričić,  Lana and L’Huillier,  Anne and Arnold,  Cord L.},
  year = {2016},
  month = dec,
  pages = {190}
}

@article{Loriot2013,
  title = {Self-referenced characterization of femtosecond laser pulses by chirp scan},
  volume = {21},
  ISSN = {1094-4087},
  url = {http://dx.doi.org/10.1364/OE.21.024879},
  DOI = {10.1364/oe.21.024879},
  number = {21},
  journal = {Optics Express},
  publisher = {Optica Publishing Group},
  author = {Loriot,  Vincent and Gitzinger,  Gregory and Forget,  Nicolas},
  year = {2013},
  month = oct,
  pages = {24879}
}

@article{Comin2014,
  title = {Compression of ultrashort laser pulses via gated multiphoton intrapulse interference phase scans},
  volume = {31},
  ISSN = {1520-8540},
  url = {http://dx.doi.org/10.1364/JOSAB.31.001118},
  DOI = {10.1364/josab.31.001118},
  number = {5},
  journal = {Journal of the Optical Society of America B},
  publisher = {Optica Publishing Group},
  author = {Comin,  Alberto and Ciesielski,  Richard and Piredda,  Giovanni and Donkers,  Kevin and Hartschuh,  Achim},
  year = {2014},
  month = apr,
  pages = {1118}
}

@article{Alonso2020,
  title = {Compact in-line temporal measurement of laser pulses with amplitude swing},
  volume = {28},
  ISSN = {1094-4087},
  url = {http://dx.doi.org/10.1364/OE.386321},
  DOI = {10.1364/oe.386321},
  number = {10},
  journal = {Optics Express},
  publisher = {Optica Publishing Group},
  author = {Alonso,  Benjam{\'\i}n and Holgado,  Warein and Sola,  {\'I}{\~n}igo J.},
  year = {2020},
  month = may,
  pages = {15625}
}

@article{Accanto2014,
  title = {Phase control of femtosecond pulses on the nanoscale using second harmonic nanoparticles},
  volume = {3},
  ISSN = {2047-7538},
  url = {http://dx.doi.org/10.1038/lsa.2014.24},
  DOI = {10.1038/lsa.2014.24},
  number = {1},
  journal = {Light: Science \& Applications},
  publisher = {Springer Science and Business Media LLC},
  author = {Accanto,  Nicolò and Nieder,  Jana B and Piatkowski,  Lukasz and Castro-Lopez,  Marta and Pastorelli,  Francesco and Brinks,  Daan and van Hulst,  Niek F},
  year = {2014},
  month = jan,
  pages = {e143–e143}
}

@article{Davis2018,
  title = {Measuring the Single-Photon Temporal-Spectral Wave Function},
  volume = {121},
  ISSN = {1079-7114},
  url = {http://dx.doi.org/10.1103/PhysRevLett.121.083602},
  DOI = {10.1103/physrevlett.121.083602},
  number = {8},
  journal = {Physical Review Letters},
  publisher = {American Physical Society (APS)},
  author = {Davis,  Alex O. C. and Thiel,  Valérian and Karpiński,  Michał and Smith,  Brian J.},
  year = {2018},
  month = aug 
}

@article{Gamouras2013,
  title = {Simultaneous Deterministic Control of Distant Qubits in Two Semiconductor Quantum Dots},
  volume = {13},
  ISSN = {1530-6992},
  url = {http://dx.doi.org/10.1021/nl4018176},
  DOI = {10.1021/nl4018176},
  number = {10},
  journal = {Nano Letters},
  publisher = {American Chemical Society (ACS)},
  author = {Gamouras,  A. and Mathew,  R. and Freisem,  S. and Deppe,  D. G. and Hall,  K. C.},
  year = {2013},
  month = sep,
  pages = {4666–4670}
}

@article{Fittinghoff1996,
  title = {Measurement of the intensity and phase of ultraweak,  ultrashort laser pulses},
  volume = {21},
  ISSN = {1539-4794},
  url = {http://dx.doi.org/10.1364/OL.21.000884},
  DOI = {10.1364/ol.21.000884},
  number = {12},
  journal = {Optics Letters},
  publisher = {Optica Publishing Group},
  author = {Fittinghoff,  David N. and Walmsley,  Ian A. and Bowie,  Jason L. and Sweetser,  John N. and Jennings,  Richard T. and Krumb\"{u}gel,  Marco A. and DeLong,  Kenneth W. and Trebino,  Rick},
  year = {1996},
  month = jun,
  pages = {884}
}

@article{Bowlan2007,
  title = {Directly measuring the spatio-temporal electric field of focusing ultrashort pulses},
  volume = {15},
  ISSN = {1094-4087},
  url = {http://dx.doi.org/10.1364/OE.15.010219},
  DOI = {10.1364/oe.15.010219},
  number = {16},
  journal = {Optics Express},
  publisher = {Optica Publishing Group},
  author = {Bowlan,  Pamela and Gabolde,  Pablo and Trebino,  Rick},
  year = {2007},
  month = jul,
  pages = {10219}
}
\end{document}